# Apuntes relevantes sobre la evaluación en la alfabetización informacional

Lic. Carlos Luis González-Valiente*

## A MODO DE INTRODUCCIÓN

La evaluación en el contexto de la alfabetización en información (ALFIN) tiene lugar como una de las etapas que conforman el ciclo de vida de este tipo de proyectos. Aunque ella no debe verse como una etapa más, sino como un proceso de carácter iterativo que está inmerso en cada fase que conforma los programas pertinentes. Al respecto, Horton (2007) declara que a las actividades evaluativas en la ALFIN no se les presta mucha importancia; aunque ella es también parte de la medición del aprendizaje de los estudiantes[1].

Oportunamente Gratch (2006) distingue tres posibles escenarios para la evaluación de la ALFIN: el *entorno de aprendizaje*, los *componentes del programa ALFIN* y los *resultados del aprendizaje en el alumno*. Gratch afirma además que sobre tales escenarios debe hacerse un uso de múltiples medidas y métodos para el acopio de evidencias[2].

Estos ámbitos de evaluación guardan una estrecha relación con los planteamientos definidos por las teorías emanadas desde la pedagogía, para el tratamiento del fenómeno del aprendizaje. Tales teorías son la *conductual*, la *cognitiva*, la *humanística*, y la *social/institucional*[3]. Desde ellas, los aspectos evaluativos también hacen una incidencia directa sobre los estilos de aprendizaje, los cuales fijan enfoques cognitivos en los individuos permitiéndoles entender e integrar la información.

Precisamente los escenarios de evaluación ofrecen un vínculo directo con las tres fases para evaluar el aprendizaje en la ALFIN que propone Licea (2007). Estas son: (1) la *evaluación diagnóstica*, para conocer el grado de conocimientos previos de los participantes en actividades de estudio, (2) las *evaluación formativa*, para conocer las fortalezas o debilidades de la ALFIN y (3) la *evaluación sumaria*, para identificar las medidas que deben tomarse en relación con la continuidad o la suspensión del programa o el logro de sus objetivos y metas. A su vez, la autora afirma que "no es posible mejorar la enseñanza sin conocer su efecto". Por lo que, los resultados de la actividad evaluativa pueden servir para diagnosticar la ALFIN en sí misma, a los instructores, a los participantes actuales y los que ya participaron; así como medir el progreso de éstos [4]. Es por ello que la aplicación de medios para el diagnóstico es un modo para la evolución de la ALFIN, la cual apunta a la valoración continua, progresiva y sistemática de los componentes que la integran.

Por otra parte McCulley (2009) sostiene que "una valoración exitosa incluye mediciones sumativas y formativas, ambas como una parte integral del proceso de aprendizaje"[5]; donde el centro de atención debe estar encaminado hacia la enseñanza sobre los estudiantes y no en la puntuación de los exámenes, ya que esto constituye un proceso recursivo, donde los resultados pueden ser usados para mejorar la enseñanza, lo cual también se convierte en un factor crítico para la mejora del aprendizaje.

Licea (2007) concibe que la evaluación de la ALFIN utiliza métodos que se comparten con otras disciplinas y citando a Kirby *et al.* (1998) distingue algunos ejemplos de ellos como: la observación, la retroalimentación de los usuarios, los cuestionarios in situ, los deberes, el monitoreo, la evaluación por colegas y las reflexiones personales, entre otros[4]. Aunque, es válido acotar que "una buena práctica de evaluación adquiere y admite el uso de variados métodos para el acopio de evidencias"[2].

Azura *et al.* (2008) afirman que "la alfabetización informacional debería hacer un esfuerzo para experimentar con diferentes enfoques para encontrar uno, o la combinación de algunos métodos, que mejor se adapten a la clase que enseñan sobre la base de los estilos de aprendizaje y preferencias de los estudiantes"[3]. Por lo que otra cuestión importante en la evaluación es el contexto, lo cual ha sido un tema abordado en la literatura especializada por LLoyd y Williamson (2008)[6]. El contexto incide en las actitudes metodológicas que adopta la ALFIN. Esto es un elemento clave porque ayuda a ésta desde una perspectiva práctica y desde sus procesos para facilitar el aprendizaje. Crawford e Irwing (2009) distinguen que "el conocimiento se forma por el contexto, éste se incrusta en una situación que comprende una ubicación, un conjunto de actividades y un conjunto de relaciones sociales en las que estas actividades son colocadas". Estos autores también señalan como una discusión emergente lo relativo al concepto del modelo del estudiante múltiple, es decir, las mismas personas aprendiendo en diferentes contextos: el personal, la comunidad, y el lugar de trabajo[7].

## IMPLICACIONES PROFESIONALES

Los profesionales de la información (PI), como diseñadores y ejecutores de la ALFIN, deben orientarse hacia su evaluación; porque más allá de que ésta indica el posible cierre del ciclo, es el proceso que garantiza la comprobación directa de cuánto se ha evolucionado respecto a ideas, conocimientos y experiencias relativas al uso de la información. Gratch (2006) advierte que una de las cuestiones por las que estos profesionales obvian la evaluación es por el bajo nivel de capacitación y experiencias en metodologías de evaluación y análisis de datos, por el carácter complejo y multidimensional del aprendizaje y por cuestiones relativas al uso de recursos humanos, financieros y de tiempo[2].

El PI no solo requiere tener un dominio sobre los contenidos a impartir en un programa o de la manera en que los va a diseñar, sino que necesita tener un control y un acercamiento directo con el ambiente de aprendizaje de sus usuarios; para así incidir y negociar con los métodos de aprendizaje formal e informal, de manera tal que le permita adoptar formas de evaluación adecuadas. Aunque la evaluación no está limitada a uno o pocos modos de aplicación, sino que es multivariada y está en constante evolución. La manera en que el PI logre dominar el "entorno de aprendizaje" se lograrán aplicar eficaces técnicas, instrumentos, recursos y métodos para ello. La actualización de los contenidos es otra cuestión, al igual que las normas y herramientas que brinda la ALFIN. Gratch (2006) afirma que: "no es necesario convertirse en un experto en todo; en realidad, saber qué tipo de cuestiones hay que afrontar y dónde encontrar más información y ayuda, es el primer requisito para llevar a cabo estudios de evaluación"[2].

## BIBLIOGRAFÍA

[*] Especialista en información del Departamento de Informática y Gestión de la Información del Grupo Empresarial de la Industria Sidero Mecánica (GESIME), carlos.valiente@fcom.uh.cu; cvaliente@sime.cu